\documentclass[A4,conference]{IEEEtran}
\usepackage[T1]{fontenc}
\usepackage[latin9]{inputenc}
\usepackage{color}
\usepackage{verbatim}
\usepackage{booktabs}
\usepackage{amsmath}
\usepackage{amsthm}
\usepackage{graphicx}
\usepackage[detect-all,detect-weight=true]{siunitx}
\usepackage{xcolor}
\usepackage{multirow}
\usepackage{colortbl}
\usepackage{cite}
\usepackage{array}
\usepackage{url}

\makeatletter

\providecommand{\tabularnewline}{\\}

\definecolor{darkgreen}{RGB}{11,102,35}
\theoremstyle{plain}
\newtheorem{thm}{\protect\theoremname}
\theoremstyle{remark}
\newtheorem{rem}[thm]{\protect\remarkname}
\theoremstyle{plain}
\newtheorem{defn}{Definition}

\def\endthebibliography{%
	\def\@noitemerr{\@latex@warning{Empty `thebibliography' environment}}%
	\endlist
}

\makeatother

\providecommand{\remarkname}{Remark}
\providecommand{\theoremname}{Theorem}

\DeclareSIUnit\decibelSM{dBsm}

\begin{document}
\sisetup{range-phrase=\text{--},range-units=single}

\title{Radar Communication for Combating Mutual Interference of FMCW Radars}
\author{Canan Aydogdu, Nil Garcia, Lars Hammarstrand, and Henk Wymeersch\\
Department of Electrical Engineering, \\ Chalmers University of Technology,
Sweden\\
e-mail: canan@chalmers.se }
\maketitle

\begin{abstract}
Commercial automotive radars used today are based on frequency modulated
continuous wave signals due to the simple and robust detection
method and good accuracy. However, the increase in both the number
of radars deployed per vehicle and the number of such vehicles leads
to mutual interference, cutting short
future plans for autonomous driving and active safety functionality. We propose and analyze
a radar communication (RadCom) approach to reduce this
mutual interference while simultaneously offering communication functionality.
We achieve this by frequency division multiplexing radar and communication,
where communication is built on a decentralized carrier sense multiple
access protocol and is used to adjust the timing of radar transmissions.
Our simulation results indicate that radar interference can be significantly
reduced, at no cost in radar accuracy.
\end{abstract}

\section{Introduction}

\begin{figure}
\centering{}\includegraphics[clip,width=0.9\columnwidth]{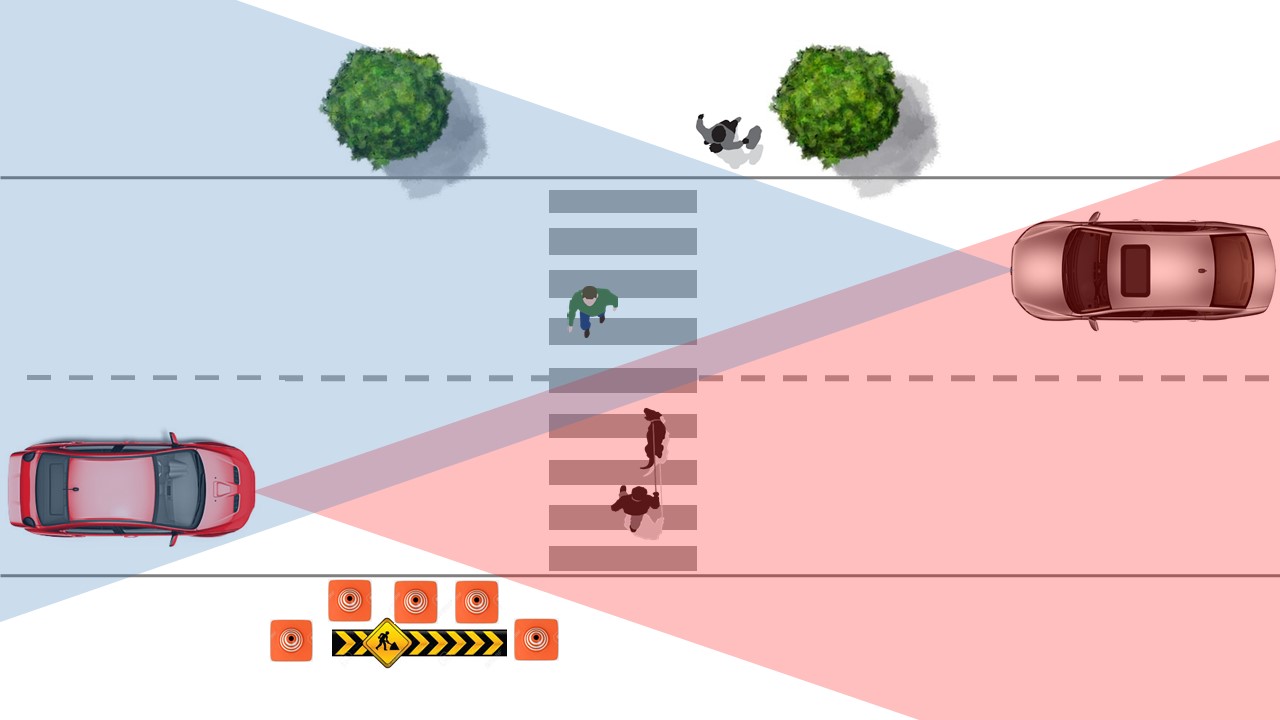}
\caption{Scenario with two vehicles in each other's field of view, leading
to mutual interference.}
\label{fig_scenario}
\end{figure}

Automotive radar is becoming an indispensable equipment in modern cars,
for different functions including adaptive cruise control and
parking, especially due to its insensitivity to bad weather conditions~\cite{patole2017automotive}.
Today, most automotive radar systems operate at 76\textendash 81 GHz
\cite{jurgen2012}, which provides good range resolution, on the
order of centimeters\cite{patole2017automotive} and the possibility to
mitigate interference by locating the radars at different carrier
frequencies \cite{schipper2012systematic}.
Likewise, vehicle-to-vehicle (V2V) communication is becoming a standard, having proven its value in dissemination of safety
critical information~\cite{patole2017automotive,kong2017}. However,
both technologies have limitations related to the increased
penetration rates. For example, current automotive radar sensors are not controllable
or able to coordinate with sensors on other vehicles. Hence, mutual
radar interference becomes a problem as in Fig.~\ref{fig_scenario}, resulting in increased noise
floor \cite{goppelt2011analytical,bourdoux2017}, in turn
resulting in reduced detection capability and ghost detections \cite{MosarimFinalReport}. 

Radar
Communication (RadCom) is an approach to use radar hardware for communication
purposes and can rely on the highly under-utilized frequency modulated
continuous wave (FMCW) radar bandwidth.
This idea was first applied in a vehicular application, where radar
and communication signals are transmitted between a vehicle and a road
side unit with a spread spectrum technique by two different chipping
sequences~\cite{takeda1998}.
Most of the other works consider orthogonal frequency
division multiplexing (OFDM) for radar communications~\cite{donnet2006,garmatyuk2011,sturm2011}.
OFDM is widely used for communication due to its high degree of flexibility,
low receiver complexity, and high performance under different propagation
conditions. The signal processing aspects of OFDM based RadCom is
identified with its potential optimizations in~\cite{falcone2010,sturm2010,reichardt2012}
together with experimental test measurements to prove its applicability.
An important
challenge is that standard communication frequencies, such as the
5.9 GHz ISM band are generally unsuitable for radar, as the achieved
accuracy and resolution is insufficient to meet automotive requirements~\cite{reichardt2012}.
In higher frequencies, a preliminary study for joint mmWave RadCom
for a vehicular environment in the 60GHz band~\cite{kumari2015}, used
the IEEE 802.11ad preamble as a radar signal with standard WiFi receiver
algorithms. This technique is shown to achieve a reasonable range and velocity estimation
accuracy (0.1m and 0.1m/s). A similar
study demonstrated how automotive sensor data 
    is
beneficial to the mmWave beam alignment, pointing out the possibility
of adding mmWave communication functions on existing mmWave automotive
radars in the 76-81 GHz band~\cite{junil2016}. 

In this paper, we propose and analyze a novel RadCom approach for
high-frequency FMCW radars, in which we repurpose a small part of the
radar bandwidth to create an 802.11p-like V2V connection. The V2V
channel is controlled via a carrier sense multiple access (CSMA) protocol
and is utilized to control the timing of the radar signals. Vehicles are assumed to perform \si{\micro\second}-level clock synchronization with GNSS. We have performed
an in-depth simulation of the proposed concept for a two-vehicle scenario.
We find that under realistic propagation conditions, RadCom can significantly
reduce the radar interference, with no performance
degradation in terms of radar accuracy.

\section{System Model}

\label{sec_model}

\label{sec_radar}

\subsection{FMCW Transmitter}
\begin{figure}
\centering{}\includegraphics[width=\linewidth]{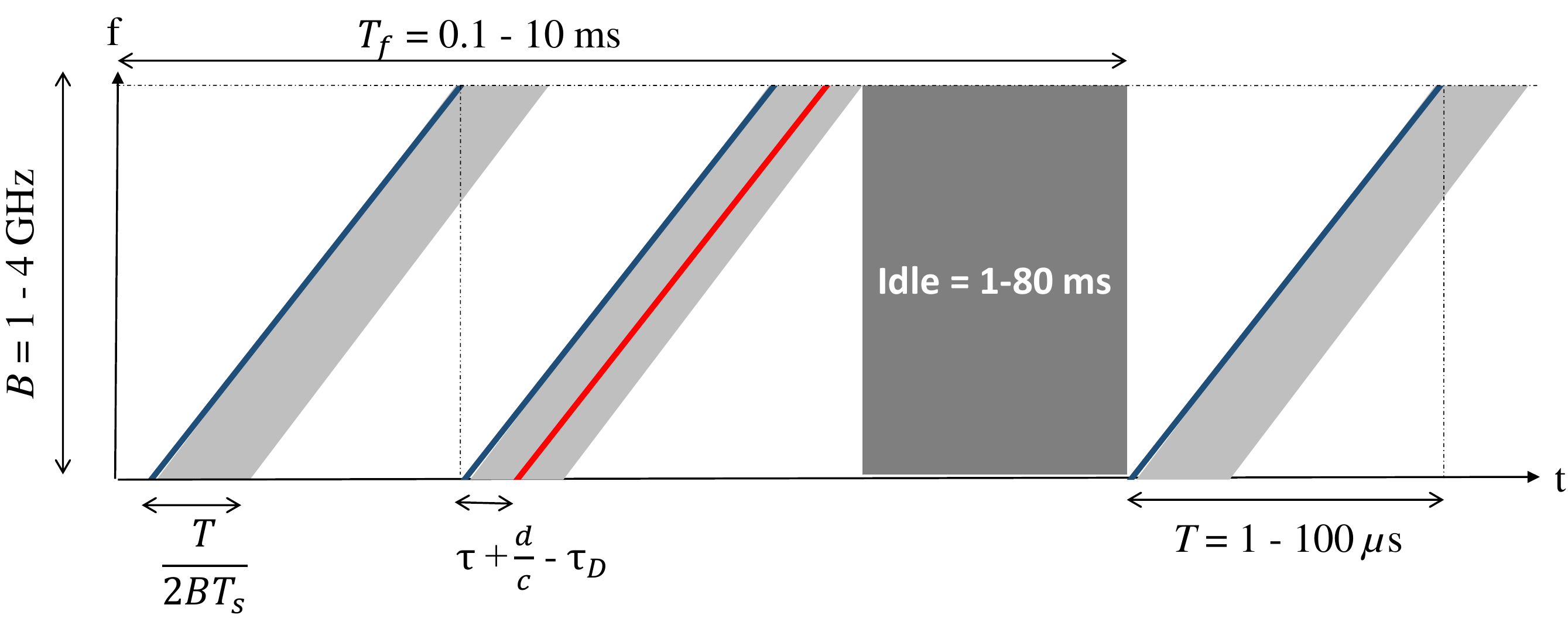}
\caption{FMCW modulation used in most automotive radar, with typical operating parameters. For a single radar, much of the time-frequency space is not utilized (the idle time is not shown to scale). The red line represents an interfering radar signal from another vehicle.}
\label{fig:FMCW}
\end{figure}

We consider a sequence of frequency modulated continuous waves, i.e.,
chirps, transmitted by an FMCW radar, of the form
\begin{align}
& s(t)=\sqrt{P_{t\text{x}}}\sum_{k=1}^{N}c(t-kT)\label{eq:TXwaveform}
\end{align}
where $c(t)$ is a chirp of the form $c(t)=\exp\left(j2\pi\left(f_{c}+{B}t/T\right)t\right)$, 
$P_{\text{tx}}$ is the transmit power, $B$ denotes the
radar bandwidth (typically \SIrange{1}{4}{\giga\hertz}), $f_{c}$ is the carrier frequency (\SI{77}{\giga\hertz}), $T$ is the chirp duration,
and $N$ is the number of chirps per frame. The frame time $T_{f}$ comprises $NT$ plus the idle and processing time. Depending on the maximum detectable range ($d_\mathrm{max}$) and  maximum detectable relative velocity ($v_\mathrm{max}$), plus range and velocity resolution of the FMCW automotive radar, $T$, $T_f$ and $N$ take typical values seen in Fig.~\ref{fig:FMCW}. In Fig.~\ref{fig:FMCW} the signal format (frequency occupancy as a function of time) of a sawtooth FMCW signal is shown, where the blue line is the transmitted chirp, the grey band corresponds to the sampling bandwidth of the receiver and the red line is an interference received from a FMCW radar which started transmission with a $\tau$ time difference.
After $N$ successive chirps, there is a significant idle time used for processing the samples. Further, the white region indicates
a large fraction of unused time-frequency resources.

\subsection{FMCW Receiver}
At the co-located receiver, the backscattered signal is processed.
The radar receiver comprises of the following blocks \cite{refOfMatlab}:
a mixer, an analog-to-digital convertor (ADC), and a digital processor.
The mixer multiplies the received signal with a copy of the transmitted
chirp. After low-pass filtering the resulting intermediate frequency
(IF) signal, the mixer will output a signal with multiple harmonics
at frequencies proportional to the time difference between the transmitted
chirp and the received chirps. The output of the mixer is then sampled
by the ADC, with sampling interval $T_{s}$, and passed to the digital
processor which will detect and estimate the frequencies. The ADC
bandwidth $1/(2 T_{s})$ is generally on the order of \SIrange{10}{50}{\mega\hertz} and is
thus much smaller than $B$. Considering a single target at distance $d$,  the sampled back-scatter signal, sample $n$ or
chirp $k$ is of the form \cite{patole2017automotive}
\begin{equation}
r_{n}^{(k)}=\sqrt{\gamma P_{\text{tx}}d^{-4}}\exp\left(j2\pi\frac{B(2d/c -2\tau_{D}) }{T}nT_{s}\right)+w_{n}^{(k)}\label{eq:RXwaveform}
\end{equation}
where $\gamma= G_{\mathrm{tx}}G_{\mathrm{rx}}\sigma \lambda^2/(4\pi)^3$, for  target radar cross section (RCS) $\sigma$, transmitter and receiver antenna gains $G_{\mathrm{tx}}$ and $G_{\mathrm{rx}}$, Doppler time shift $\tau_{D}={Tvf_c}/{(Bc)}$, in which $c$ denotes the speed of light, $v$ is the relative velocity between vehicles (note that a positive $v$ corresponds to approaching vehicles and a positive Doppler shift, which leads to a decreased time difference between the transmitted and reflected radar signal), $w_{n}^{(k)}$ is additive white Gaussian noise (AWGN) with variance $N_{0}$.  A common approach to frequency retrieval in FMCW radar is to compute the fast Fourier transform (FFT) of the signal, average the signal through multiple chirp periods for enhanced SNR, and detect the peaks in the frequency-domain.

\subsection{Goal}
Our aim is to study the performance of the receiver when the target
is itself a vehicle with an FMCW radar. We propose an FMCW-based RadCom system and  investigate how the probability of mutual interference, the probability of false
alarm and the ranging error are effected by the proposed scheme. This study focuses on resolving radar conflicts among vehicles with similar radar characteristics (same bandwidth and chirp signal), which is easier to analyze than the more general heterogeneous scenario.

\section{Radar Interference Analysis}

In this section, we describe the interference model and calculate
the conditions under which interference exists, both for a single
link and for a network.

\subsection{Interference Model\label{subsec:Interference-Model}}

\label{sec_theory}

\begin{figure}
\centering{}\includegraphics[width=0.95\linewidth]{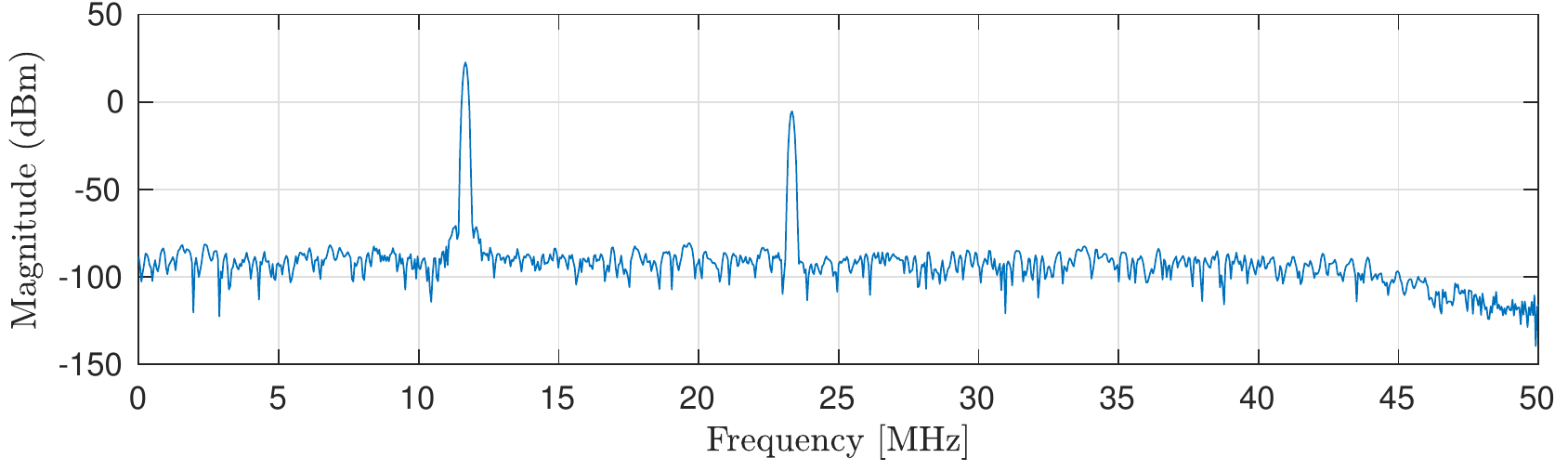}
\caption{FFT of received signal, with a interfering vehicle target at 70 m (23.33 MHz) and  a ghost target at 35 m (11.67 MHz).}
\label{figSignalWindowed}
\end{figure}

Consider the scenario in Fig.~\ref{fig_scenario}. Two cars with
front-mounted FMCW radars facing each other approach a road-crossing where
pedestrians are present. 
 Both radars are FMCW based and use the same frequency band.
 If the
interfering radar is located at distance $d$ and has a relative delay $\tau$ between the ego vehicle transmission and the interfering
vehicle transmission, then the received signal
at the ego radar becomes
\begin{align}
 & \tilde{r}_{n}^{(k)}\label{eq:interference}\\
 & =\begin{cases}
r_{n}^{(k)} & \tau \notin V\\
r_{n}^{(k)}+\sqrt{\tilde{\gamma}P_{\text{tx}}d^{-2}}\exp\left(j2\pi\frac{B(\tau+d/c- \tau_D )}{T}nT_{s}\right) & \tau \in V
\end{cases}\nonumber
\end{align}
where $\tilde{\gamma}= G_{\mathrm{tx}}G_{\mathrm{rx}} \lambda^2/(4\pi)^2$
and $V$ is the \textit{vulnerable period} defined below.
\begin{defn}[Vulnerable period $V$]
Given an ego vehicle radar that starts an FMCW transmission at time $t=0$ and a facing vehicle radar with overlapping field-of-view that starts a transmission at time $t=\tau$, the vulnerable period $V$ is the set of $\tau$ values for which interference to the ego vehicle radar occurs. 
\end{defn}
As illustrated in Fig.~\ref{fig:FMCW}, the interfering signal transmitted by a time difference of $\tau$ is received at the ego radar at $\tau +d/c-\tau_{D}$. We note that interference has a factor $d^{-2}$ while the useful signal has a factor $d^{-4}$, leading to an interference that is much stronger than the useful signal. Fig.~\ref{figSignalWindowed} shows an example
of such a received FMCW signal, for a vehicle target at distance 70 m from the radar, with $\tau = 0$, which in this case is in the vulnerable period. The target is observed at 23.33 MHz, whereas the strong interference is observed at 11.67 MHz, corresponding to 35 m. 

\subsection{Single Link Interference Condition\label{subsec:Single-Link-Interference}}
In this section, we will quantify the conditions under which interference occurs. 
The interfering transmission arrives at the ego vehicle with a propagation delay $t_\text{prop}$ and a Doppler frequency shift of $f_D$. The Doppler shift is perceived by the ego vehicle as a delay $\tau_D=f_DT/B$ (see Fig.~\ref{fig:FMCW}).
Hence, the first received chirp at the ego vehicle starts at $t^\prime=\tau+t_\text{prop}+\tau_D$ and interference
will occur when $t'\in\left[0,T/(2BT_{s})\right]$. If we further account for imperfect
low-pass filtering at the ADC, which also causes mutual interference especially when the interference power is high, 
then interference will also occur when $t'\in\left[-T/(BT_{s}),-T/(2BT_{s})\right]$. 
The maximum propagation delay of a received radar signal (not filtered out at the radar ADC) is $2d_{\text{max}}/c$, where $d_{\text{max}}$ is the maximum detectable radar range and is related to chirp parameters by $d_{\text{max}}=cT/(4BT_s)$~\cite{skolnik}. Hence, the maximum propagation delay is $T/(2BT_{s})$. 
The Doppler shift of an interfering signal is $\tau_D=\pm Tv_{\text{max}}f_c/(Bc)=\pm 1/(4B)$ (approaching or receding), assuming that the maximum relative velocity of an interfering vehicle is equal to the maximum radar detectable relative velocity, given by $v_{\text{max}}=c/(4f_cT)$~\cite{skolnik}. Hence, the vulnerable period becomes
 \begin{equation}
V =\left[-\frac{3T}{2BT_{s}}-\frac{1}{4B},\frac{T}{2BT_{s}}+\frac{1}{4B}\right] \approx \label{eq:Vulnerable2}
 \left[-\frac{3T}{2BT_{s}},\frac{T}{2BT_{s}}\right]
\end{equation} 
and is approximated due to $\tau_D \ll t_\text{prop}$ and $T_s \ll T$, resulting with approximated duration
\begin{equation}
\lvert V \rvert  \approx\frac{2T}{BT_{s}}.
\label{eq:VulnerableDuration}
\end{equation}
From this, it follows that the probability of interference
for a single chirp is 
\begin{align}
\label{eq:chirpInt}
P_{\text{int}}^{(c)}\approx\frac{\lvert V \rvert}{T}\approx\frac{2}{BT_{s}}.
\end{align}
However, a radar transmits during a fraction $NT/T_{f}$ of the frame period and any interfering radar chirp sequence starting $(N-1)T$ prior up to the end of the radar transmission results with mutual interference due to overlapping of one or more chirps. Hence, the probability of interference for a frame for two facing radars is reduced to
\begin{equation}
P_{\text{int}}^{(f)}=
\frac{(2N-1)T\,P_{\text{int}}^{(c)}}{T_{f}}\approx
\frac{2NTP_{\text{int}}^{(c)}}{T_{f}},  \label{eq:frameInt}
\end{equation}
since for a typical automotive radar $N$ is large.

\begin{rem}
Note that we have not accounted for reflections or additional targets in the environment. Such targets could also lead to interference.
For that reason the vulnerable period could further be extended, following the same procedure as above. Additionally, if the radar has access to in-phase and quadrature samples, the vulnerable period can be reduced to $V \approx \left[-{T}/(2BT_{s}),{T}/(2BT_{s})\right]$.
\end{rem}

\subsection{Network Interference Condition}

For a network of vehicles, we first study a star topology around the
ego vehicle. When there are $M$ interfering vehicles, the probability
of interference is $P_{\text{int}}^{(M)}=1-(1-P_{\text{int}}^{(f)})^{M}.$
For a more general topology with $M$ vehicles, let $\mathcal{G}$ be a directed
radar graph with $\mathcal{G}=(\mathcal{V},\mathcal{E})$, where each
vertex corresponds to a radar $r_{i}\in\mathcal{V}$ and each edge
$e_{ij} \in \mathcal{E}$ means that radar $i$ is the field of view of radar $j$.
The average probability of interference is then given by 
    \begin{equation}
\bar{P}=\frac{1}{|\mathcal{V}|}\sum_{r_{i}\in\mathcal{V}}P_{\text{int}}^{(M_{i})}\label{eq:averageNetworkInterference}
\end{equation}
where $M_{i}$ denotes the number of edges into $r_{i}$.

\subsection{Performance under Radar Interference}

So-far we have only treated the probability of radar interference.
When there is interference, we know from (\ref{eq:interference})
that the interfering signal is generally stronger than the useful
signal. This affects radar performance in a number of ways: it leads
to ghost targets and an increase of the noise floor. Relevant performance
metrics are thus the probability of detection, the probability of
false alarm, and the ranging accuracy of the targets. These metrics
will be analyzed in detail in Section \ref{sec_results}.

\section{RadCom for Interference Reduction}

An FMCW-based-RadCom system is composed of three parts for sharing
the wireless channel resource:
\begin{enumerate}
\item \textit{Multiplexing} scheme for sharing among radar and communication,
\item \textit{Radar MAC} scheme (rMAC) for coordination of radar sensing
among different vehicles and
\item \textit{Communication MAC} scheme (cMAC) for sharing among different
vehicles.
\end{enumerate}
In this study, we propose a RadCom scheme, where radar and communication are frequency division multiplexed (FDM)\cite{CohabitationCOM} with time division multiple access for radar signals (denoted rTDMA) and carrier-sense multiple access for communication signals (denoted cCSMA). The proposed RadCom scheme uses communication in order to: (1) Disseminate non-overlapping rTDMA slots among radars to mitigate interference and (2) communicate data utilizing radar idle times. The focus of this article is the first function\footnote{The latter function is investigated in future studies and fully exploits FDM, whereas realization of the first function is based on frequency division.}.     

\subsection{Multiplexing}

Although multiplexing can be avoided by considering a joint waveform
for communication and radar \cite{sturm2010}, such an approach is not suitable for
automotive applications due to the limited ADC capabilities, which
do not support full-band communication (e.g., OFDM) and modulating
FMCW chirps would lead to extremely low data rates. Hence, we divide
up the bandwidth $B$ into a radar band $B_{r}$ and a communication
bandwidth $B_{c}$, for which $B_{r}+B_{c}\le B$ and $B_{c}<1/2T_{s}$,
in order to be able to reuse the radar ADC.

\subsection{Radar MAC}


\begin{figure}
\centering{}\includegraphics[width=1\columnwidth]{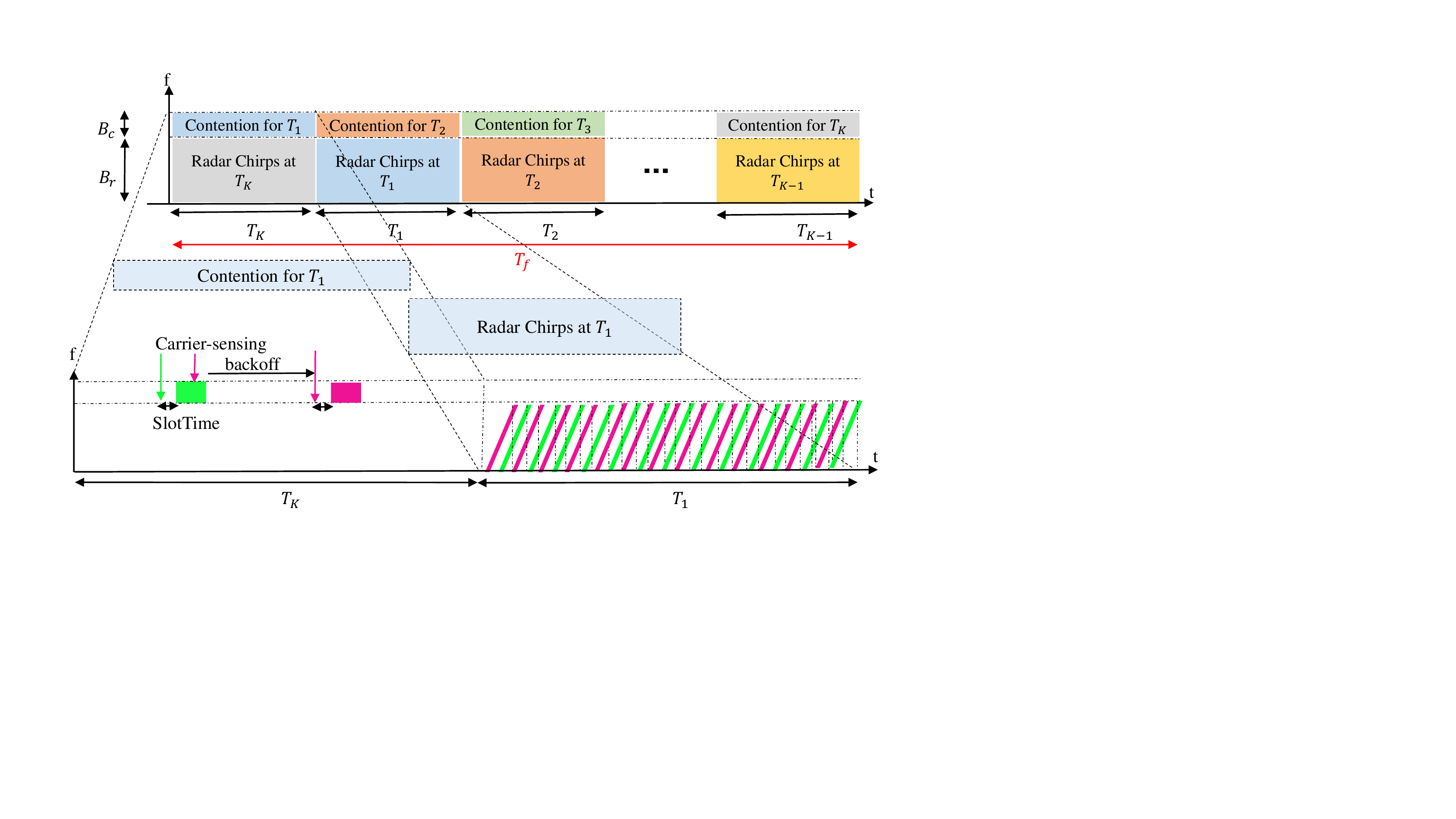}
\caption{RadCom scheme: FDM / rTDMA / cCSMA.  }
\label{figConf}
\end{figure}

Facing FMCW radars operate interference-free if they are assigned non-overlapping rTDMA slots, which depend on the vulnerable duration. For this, vehicles are assumed to synchronize their clocks using GPS. Fig.~\ref{figConf} illustrates the division of the frequency-time domain for the proposed FDM/rTDMA/cCSMA based RadCom system. 
One radar frame duration $T_f$ is divided into time slots $T_i$, where each radar transmits its chirp sequence during one $T_i$ and remains idle during rest of the frame. One time slot $T_i$ is of length $(N+1)T$, which corresponds to the duration for sending $N$ chirps plus one idle chirp time accounting for the overflow of time shifted rTDMA slots. This slotted time is set to provide non-overlapping chirp sequences and thereby maximize the number of vehicles with no mutual interference in the RadCom system, denoted by $M_\mathrm{max}$.
Using the duration of the vulnerable period $|V|$ derived in (\ref{eq:VulnerableDuration}), for the proposed rTDMA, at most $\lfloor {T}/{| V |} \rfloor$ different vehicle radars can coexist in a slot $T_i$ and the maximum number of time slots per frame is $K=\lfloor {T_f}/{(N+1)T} \rfloor$, which limits $M_\mathrm{max}$ under perfect communication to 
\begin{equation}
\label{eq:Mmax}
    M_\mathrm{max} \leq K \left \lfloor {T}/{| V |} \right \rfloor \approx K \left \lfloor {B_rT_s}/{2} \right \rfloor.
\end{equation}

\begin{figure}
\centering{}\includegraphics[width=0.6\columnwidth]{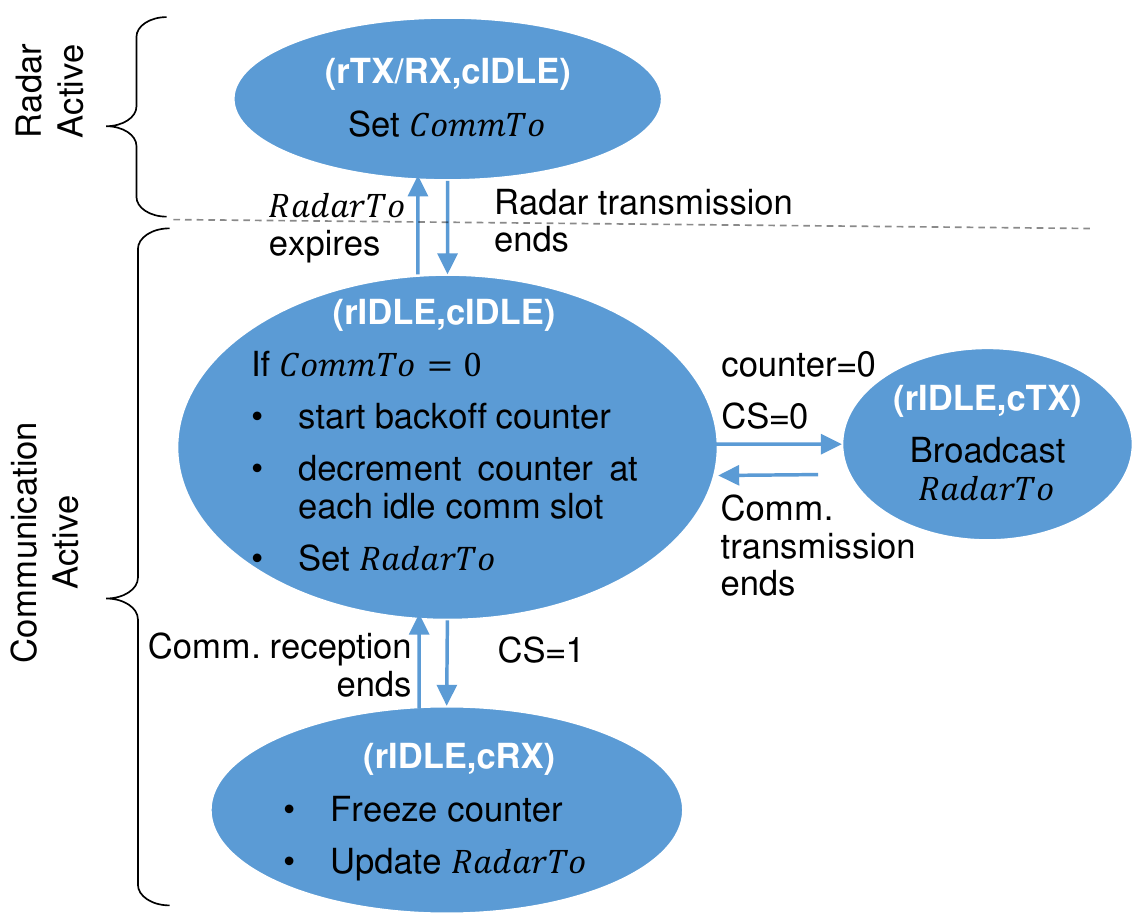}
\caption{State diagram for the proposed RadCom scheme. }
\label{fig_RadCom_StateDiagram}
\end{figure}

\subsection{Communication}
V2V vehicular communication is used to assign non-overlapping rTDMA slots among facing vehicles. Since communication links are not necessarily symmetric due to the directivity of the radar, a best effort approach with no acknowledgements is employed. Communication during slot $T_{i-1}$ determines the rTDMA slots in $T_i$ as illustrated in Fig.~\ref{figConf}.
Each time slot $T_i$ is further divided to slots called \textit{SlotTime}s, which are used by the non-persistent CSMA mechanism. Communication packets are transmitted if the channel is sensed idle for one \textit{SlotTime} or a random backoff is employed if channel is sensed busy. Each vehicle sequentially uses two timeouts: (1) Communication timeout ($CommTo$) for starting transmission of a communication packet  and (2) Radar timeout ($RadarTo$) for starting the radar chirp sequence. A communication packet (which includes the $RadarTo$ information) is transmitted upon expiration of $CommTo$, whereas the FMCW radar chirp sequence is transmitted upon expiration of $RadarTo$.  

A state diagram for the proposed RadCom scheme is given in Fig.~\ref{fig_RadCom_StateDiagram}, where each state denoted by $(rX,cY)$ corresponds to \textit{r}adar state $X$ and \textit{c}ommunication state $Y$. Radar is active in (rTX/RX,cIDLE) state and upon end of radar transmission the RadCom hardware enters (rIDLE,cIDLE) state, where communication is active and carrier sensing is deployed. Upon expiration of \textit{CommTo}, a backoff counter starts and decrements for each idle \textit{SlotTime}. If carrier is sensed idle when the counter expires ($counter=0$ and $CS=0$), the state (rIDLE,cTX) is entered, where the vehicle broadcasts its $RadarTo$. Otherwise if carrier is sensed busy, the state (rIDLE,cRX) is entered ($CS=1$), where reception takes place and any active backoff counters are frozen. In this state, the ego vehicle stores the received radar timeout values of other vehicles and updates its own $RadarTo$ to be advertised according to the received radar timeout so as to use one of the left rTDMA slots in $T_i$ or in $T_{i+1}$, etc. No measures are taken in case of communication failures and there remains a small probability for mutual interference with the RadCom scheme, which will be analyzed in a future study.

\section{Performance Evaluation and Results}
\label{sec_results}

The performance of an FMCW receiver when the target
is itself a vehicle with an FMCW radar is investigated. A comparison is made by the proposed FMCW-based RadCom system in terms of the probability of false alarm and the ranging error. The probability of mutual interference and the necessary time to resolve contention for multiple vehicles are evaluated.

\subsection{Simulation Parameters}
\label{sec_assumptions}

The simulation parameters are summarized in Table~\ref{tab:parameters}. Two facing vehicles are assumed to have radars with the same properties. Radar is FMCW with sawtooth waveform.
\begin{table}
\centering \caption{Simulation parameters.}
 %
\begin{tabular}{lll}
\toprule
 & \textbf{Parameter}  & \textbf{Value} \tabularnewline
\midrule
\multirow{5}{*}{\rotatebox{90}{Radar}}
 & Chirp duration ($T$)  & \SI{20}{\micro\second} \tabularnewline
 & Frame duration ($T_f$)  & \SI{20}{m\second} \tabularnewline
 & Time slots per frame ($K$)  & 10  \tabularnewline
 & Radar bandwidth   & \SI{0.96}{\giga\hertz}--\SI{1}{\giga\hertz}  \tabularnewline
     & $d_{\mathrm{max}}$ for $B_c=0$ & 150 m \tabularnewline
        & $v_{\mathrm{max}}$ & 140 km/h \tabularnewline
  & $P_\mathrm{tx}$  & \SI{11}{\decibel} \tabularnewline
  & $SNR$  & \SI{10}{\decibel} \tabularnewline
   & $N$  & 99 \tabularnewline
    & $f_c$ & 77 \tabularnewline
        & $T_s$ & \SI{0.01}{\micro\second} \tabularnewline
 & Chebyshev low-pass filter order  & 13 \tabularnewline
 & Thermal noise temperature  & \SI{290}{\kelvin} \tabularnewline
 & Receiver's noise figure  & \SI{4.5}{\decibel} \tabularnewline
\midrule
\multirow{5}{*}{\rotatebox{90}{Comm.}}
& Communication bandwidth $B_c$  & \SI{20}{\mega\hertz},\SI{40}{\mega\hertz}  \tabularnewline
 & Packet size ($N_{\mathrm{pkt}}$) & 4800 Bits \tabularnewline
 & Modulation  & 16-QAM \tabularnewline
 & MAC  & non-persistent CSMA 
  \tabularnewline
  & SlotTime& \SI{10}{\micro\second}
 \tabularnewline
  & Backoff window size& 6
  \tabularnewline
\bottomrule
\end{tabular}\label{tab:parameters}
\end{table}
The chirp sequence is designed so as to meet the maximum detectable relative velocity $v_{\mathrm{max}}=\SI{140}{\kilo\meter/\hour}$, the maximum detectable range $d_{\mathrm{max}}=\SI{150}{\meter}$ when $B_c=0$ (since it increases for RadCom), velocity resolution smaller than \SI{1}{\meter/\second} and range resolution of \SI{15}{\centi\meter}.
Radar front-end-hardware component parameters are taken as in~\cite{refOfMatlab}. The mean value for the radar cross section of a car is taken as \SI{20}{\decibelSM}~\cite{Lee2016,refOfMatlab}. At the signal processing stage, coherent pulse integration is applied. Moreover, a Blackman-Harris window to reduce the height of the sidelobes is applied before the FFT module. Finally, greatest of cell averaging constant false alarm rate (GoCA-CFAR) thresholding with 50 training cells with 2 guard cells is used for radar detection. The vulnerable duration for $B_c=\SI{20}{\mega\hertz}$ is computed to be $| V |=\SI{4.17}{\micro\second}$ \eqref{eq:Vulnerable2}, leading to maximum 4 concurrent radar transmissions per $T_i$, resulting with $M_\mathrm{max}=40$ vehicles supported maximum by the proposed RadCom system.

\subsection{Results}

We first consider probability of the radar interference for a multi-vehicle scenario, and then evaluate the RadCom performance in terms of delay to resolve contentions. Finally, we present in-depth results for two vehicles, with and without RadCom.

\subsubsection{Radar Interference}
\begin{figure}
\centering{}\includegraphics[width=1\columnwidth]{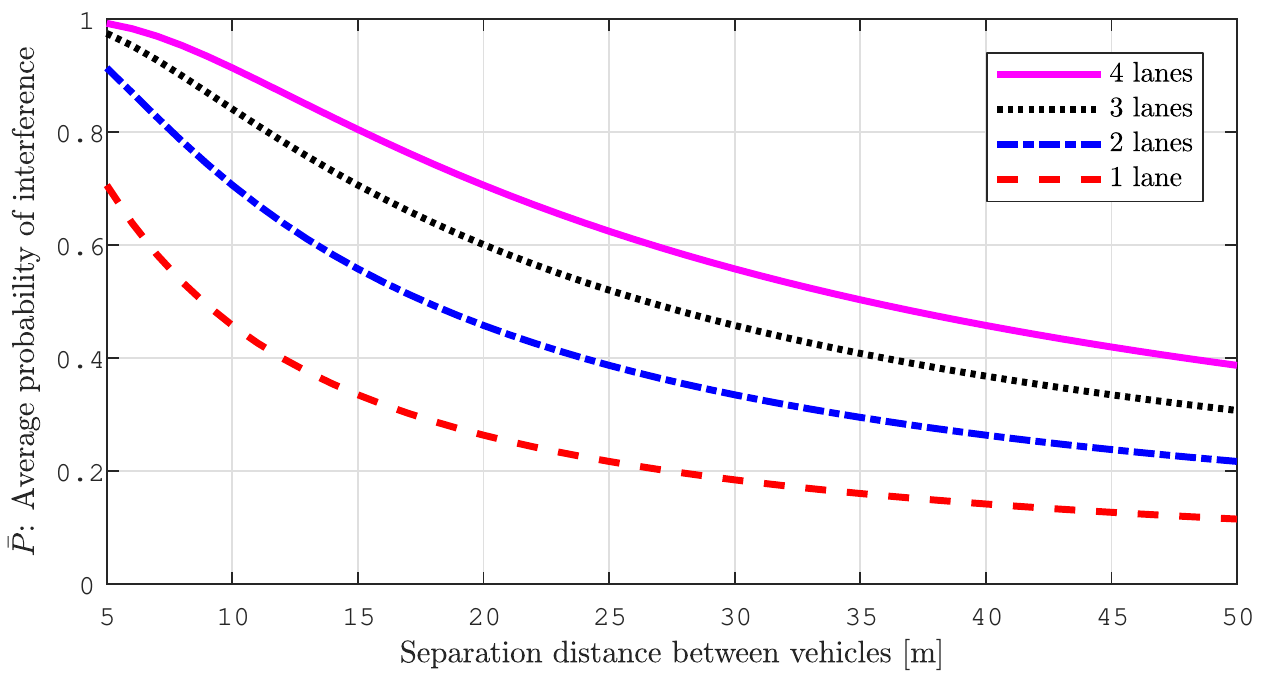}
\caption{Average probability of interference for varying vehicle separation distance and different number of lanes for $B_c=\SI{20}{\mega\hertz}$.}
\label{figPint}
\end{figure}

The average probability of interference, $\bar{P}$, for varying separation distance between vehicles on a lane and different number of lanes is given by Fig.~\ref{figPint} for $B_c=\SI{20}{\mega\hertz}$. The average probability of interference increases by lower vehicle separation and higher number of lanes, i.e., increased density of vehicles. We conclude that for dense multi-lane traffic, radar interference is a grave concern, despite the low duty cycle.

\subsubsection{Communication Delay}

\begin{figure}
\centering{}\includegraphics[width=0.9\columnwidth]{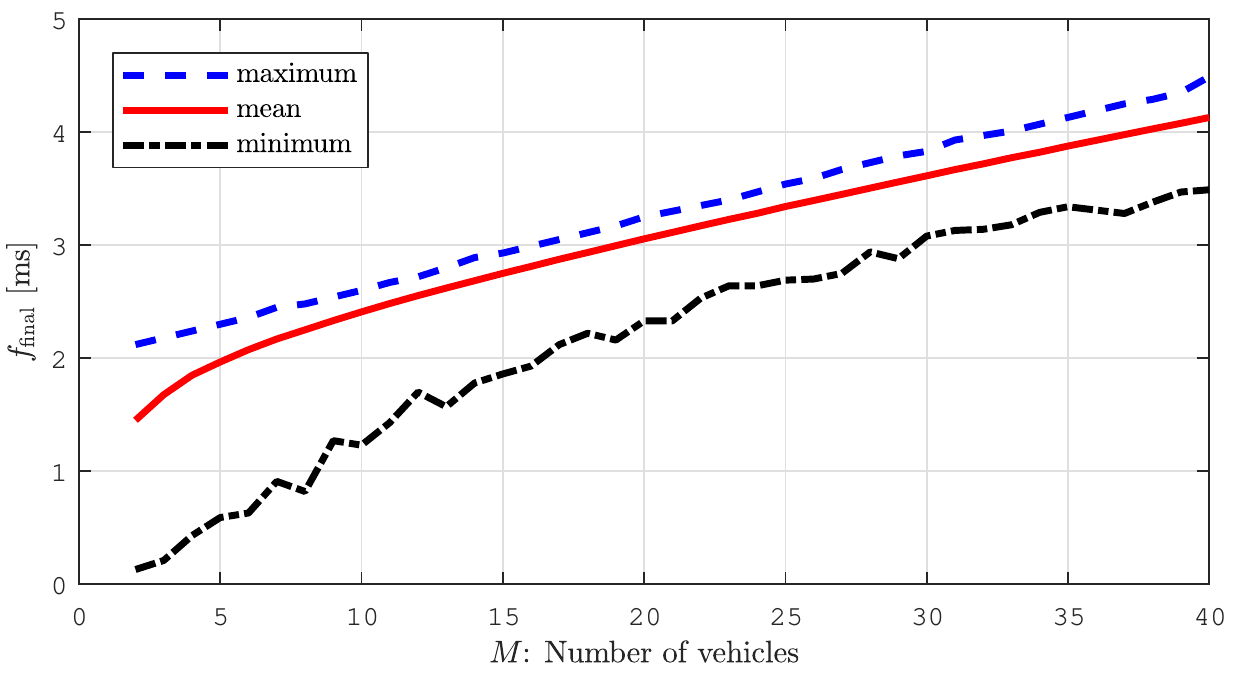}
\caption{Time necessary to resolve communication contention among $M$ vehicles for $B_c=\SI{20}{\mega\hertz}$.}
\label{figCSMA}
\end{figure}

We denote the  time to resolve contention among connected $M$ vehicles by $t_{\mathrm{final}}$. The vehicles apply non-persistent CSMA with random backoff. Based on 10,000 Monte Carlo runs, Fig.~\ref{figCSMA} shows the average, minimum, and maximum value of $t_{\mathrm{final}}$ as a function of $M$ for $B_c=\SI{20}{\mega\hertz}$. We assume that all $M$ stations initiate transmission attempts during the time slot $T_1$.
We observe that the contention time for $M_\mathrm{max}=40$ vehicles lasts up to 5 ms and is resolved within a radar frame, which is $T_f=\SI{20}{\milli\second}$. Hence, the proposed communication scheme is feasible and fast enough to allocate rTDMA slots to 40 vehicles before the radar frame ends. Furthermore, a larger communication bandwidth will lead to lower $t_{\mathrm{final}}$ leaving space for other types of communications. 
\begin{figure}
\centering{}\includegraphics[width=1\columnwidth]{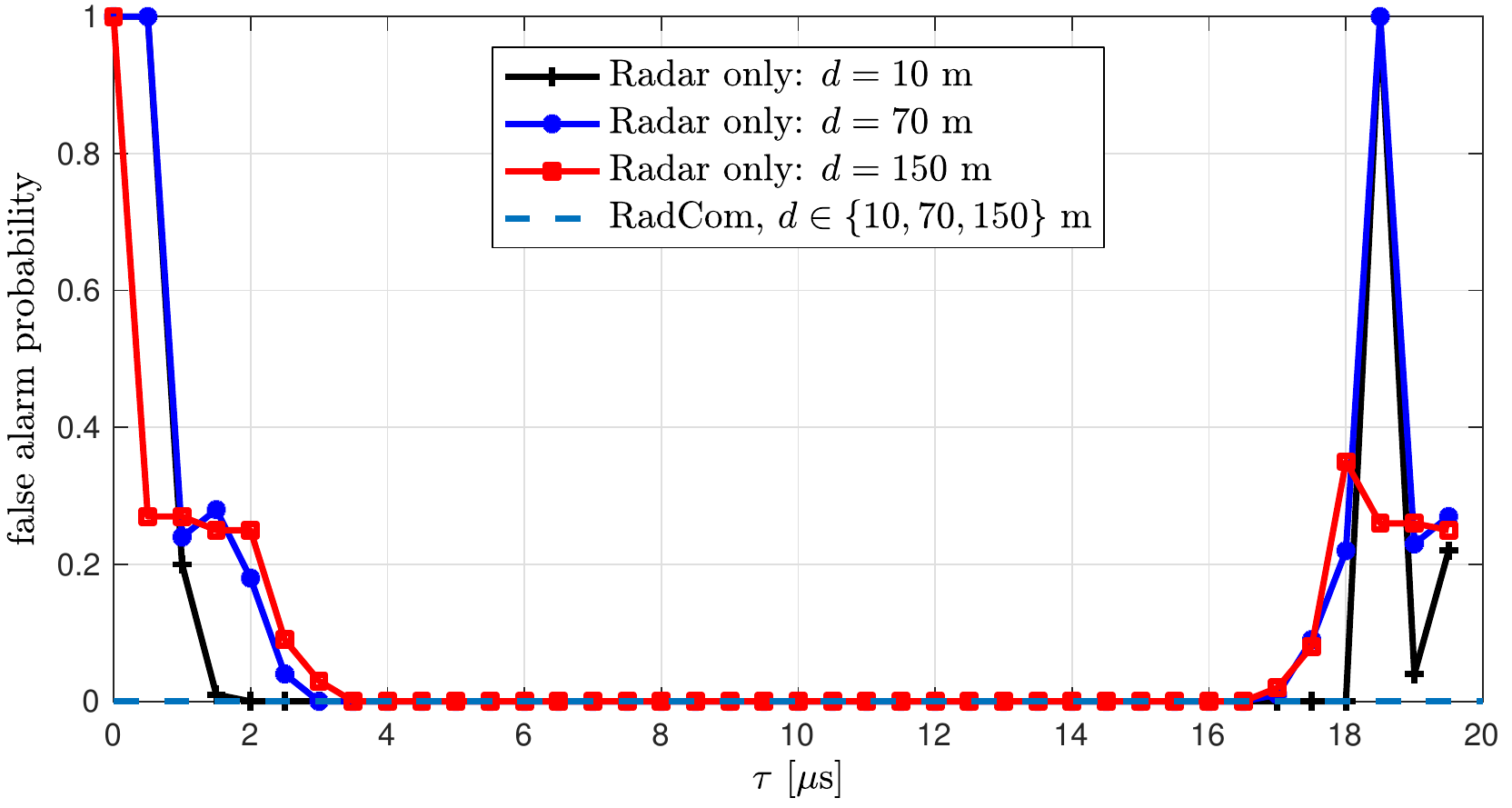}
\caption{Comparison of probability of false alarm between radar and RadCom with $B_c=\SI{20}{\mega\hertz}$.}
\label{figPfAll}
\end{figure}

\subsubsection{Radar Interference with RadCom}

The performance of RadCom is evaluated for the communication bandwidths $B_c\in \{20,40\}\si{\mega\hertz}$. The distance between two vehicles ($d$) and the time difference between
starting times of chirps ($\tau$) are varied. The performance of radar only and RadCom schemes are compared in terms of probability of false alarm ($P_f$) and ranging error based on 100 Monte Carlo simulations.
Fig.~\ref{figPfAll} shows the probability of false alarm without RadCom for $B_c=\SI{20}{\mega\hertz}$ for varying $\tau$ and three different $d$ values. The transmissions during the vulnerable period $V$ is observed to result with interference for especially $\tau< {T}/{(2BT_s)}=\SI{1}{\micro\second}$ and $\tau=\SIrange{17}{20}{\micro\second}$, described in \eqref{eq:Vulnerable2}. False alarm probability increases with increasing $d$ due to attenuated radar received signals. With RadCom, the false alarm probability is reduced to zero for all $\tau$ and $d$ due to avoiding transmissions during the vulnerable period. Note that, RadCom has non-zero probability of false alarm even with RadCom, which is not observed due to the low number of simulations.

Simulations taken over varying  $\tau$ and $d$, show that the ranging error is independent of $\tau$ and at most  \SI{6.9}{\centi\meter}, \SI{7.4}{\centi\meter} and \SI{8.54}{\centi\meter} for radar only, RadCom with $B_c=\SI{20}{\mega\hertz}$ and  $B_c=\SI{40}{\mega\hertz}$, respectively. 
While RadCom in theory can lead to an accuracy reduction due to reduced bandwidth, this effect is insignificant, since the relative decrease in radar bandwidth is negligible (at most 4\%) and we operate in high SNR.

\section{Conclusion}

We have evaluated a RadCom approach building on a combination of FDM,
TDMA for radar, and CSMA for communication. The approach exploits
the low utilization of time and frequency of a typical radar, as well
as the limited impact of a small bandwidth loss on the radar performance.
We have performed an interference analysis at both the link and network
level and found that with higher penetration, interference is prevalent.
With our proposed approach, we are able to mitigate interference by
shifting radar transmissions in time. Performance in terms of false
alarms, missed detections, and ranging accuracy are reported,
based on high-fidelity simulations. Future work will consider larger-scale scenarios for heterogeneous FMCW radars with different bandwidths and chirp parameters, as well as the interference from multipath. 

\section*{Acknowledgment}

This work is supported, in part, by Marie Curie Individual Fellowships (H2020-MSCA-IF-2016) Grant 745706 (GreenLoc)
and a SEED grant
from Electrical Engineering Department of Chalmers University of Technology.

\bibliographystyle{IEEEtran}
\bibliography{IEEEabrv,./bibliography}

\begin{thebibliography}{10}
\providecommand{\url}[1]{#1}
\csname url@samestyle\endcsname
\providecommand{\newblock}{\relax}
\providecommand{\bibinfo}[2]{#2}
\providecommand{\BIBentrySTDinterwordspacing}{\spaceskip=0pt\relax}
\providecommand{\BIBentryALTinterwordstretchfactor}{4}
\providecommand{\BIBentryALTinterwordspacing}{\spaceskip=\fontdimen2\font plus
\BIBentryALTinterwordstretchfactor\fontdimen3\font minus
  \fontdimen4\font\relax}
\providecommand{\BIBforeignlanguage}[2]{{%
\expandafter\ifx\csname l@#1\endcsname\relax
\typeout{** WARNING: IEEEtran.bst: No hyphenation pattern has been}%
\typeout{** loaded for the language `#1'. Using the pattern for}%
\typeout{** the default language instead.}%
\else
\language=\csname l@#1\endcsname
\fi
#2}}
\providecommand{\BIBdecl}{\relax}
\BIBdecl

\bibitem{patole2017automotive}
S.~M. Patole, M.~Torlak, D.~Wang, and M.~Ali, ``Automotive radars: A review of
  signal processing techniques,'' \emph{IEEE Signal Processing Magazine},
  vol.~34, no.~2, pp. 22--35, 2017.

\bibitem{jurgen2012}
J.~Hasch, E.~Topak, R.~Schnabel, T.~Zwick, R.~Weigel, and C.~Waldschmidt,
  ``Millimeter-wave technology for automotive radar sensors in the 77 {GHz}
  frequency band,'' \emph{IEEE Transactions on Microwave Theory and
  Techniques}, vol.~60, no.~3, pp. 845--860, March 2012.

\bibitem{schipper2012systematic}
T.~Schipper, M.~Harter, L.~Zwirello, T.~Mahler, and T.~Zwick, ``Systematic
  approach to investigate and counteract interference-effects in automotive
  radars,'' in \emph{{IEEE} 9th European Radar Conference}, 2012, pp. 190--193.

\bibitem{kong2017}
L.~Kong, M.~K. Khan, F.~Wu, G.~Chen, and P.~Zeng, ``Millimeter-wave wireless
  communications for {IoT}-cloud supported autonomous vehicles: Overview,
  design, and challenges,'' \emph{IEEE Communications Magazine}, vol.~55,
  no.~1, pp. 62--68, January 2017.

\bibitem{goppelt2011analytical}
M.~Goppelt, H.-L. Bl{\"o}cher, and W.~Menzel, ``Analytical investigation of
  mutual interference between automotive {FMCW} radar sensors,'' in
  \emph{GermanMicrowave Conference (GeMIC)}.\hskip 1em plus 0.5em minus
  0.4em\relax IEEE, 2011, pp. 1--4.

\bibitem{bourdoux2017}
A.~Bourdoux, K.~Parashar, and M.~Bauduin, ``Phenomenology of mutual
  interference of {FMCW} and {PMCW} automotive radars,'' in \emph{{IEEE} Radar
  Conference (RadarConf)}, May 2017, pp. 1709--1714.

\bibitem{MosarimFinalReport}
\BIBentryALTinterwordspacing
I.~M. Kunert, ``Project final report, {MOSARIM}: More safety for all by radar
  interference mitigation,'' 2012. [Online]. Available:
  \url{http://cordis.europa.eu/docs/projects/cnect/1/248231/080/deliverables/001-D611finalreportfinal.pdf}
\BIBentrySTDinterwordspacing

\bibitem{takeda1998}
M.~Takeda, T.~Terada, and R.~Kohno, ``Spread spectrum joint communication and
  ranging system using interference cancellation between a roadside and a
  vehicle,'' in \emph{IEEE 48th Vehicular Technology Conference}, vol.~3, May
  1998, pp. 1935--1939.

\bibitem{donnet2006}
B.~J. Donnet and I.~D. Longstaff, ``Combining {MIMO} radar with {OFDM}
  communications,'' in \emph{European Radar Conference}, Sept 2006, pp. 37--40.

\bibitem{garmatyuk2011}
D.~Garmatyuk, J.~Schuerger, and K.~Kauffman, ``Multifunctional software-defined
  radar sensor and data communication system,'' \emph{IEEE Sensors Journal},
  vol.~11, no.~1, pp. 99--106, Jan 2011.

\bibitem{sturm2011}
C.~Sturm and W.~Wiesbeck, ``Waveform design and signal processing aspects for
  fusion of wireless communications and radar sensing,'' \emph{Proceedings of
  the IEEE}, vol.~99, no.~7, pp. 1236--1259, July 2011.

\bibitem{falcone2010}
P.~Falcone, F.~Colone, C.~Bongioanni, and P.~Lombardo, ``Experimental results
  for {OFDM} wifi-based passive bistatic radar,'' in \emph{{IEEE} Radar
  Conference}, May 2010, pp. 516--521.

\bibitem{sturm2010}
C.~Sturm, T.~Zwick, W.~Wiesbeck, and M.~Braun, ``Performance verification of
  symbol-based {OFDM} radar processing,'' in \emph{{IEEE} Radar Conference},
  May 2010, pp. 60--63.

\bibitem{reichardt2012}
L.~Reichardt, C.~Sturm, F.~Grunhaupt, and T.~Zwick, ``Demonstrating the use of
  the {IEEE} 802.11p car-to-car communication standard for automotive radar,''
  in \emph{6th European Conference on Antennas and Propagation (EUCAP)}, March
  2012, pp. 1576--1580.

\bibitem{kumari2015}
P.~Kumari, N.~Gonzalez-Prelcic, and R.~W. Heath, ``Investigating the {IEEE}
  802.11ad standard for millimeter wave automotive radar,'' in \emph{{IEEE}
  82nd Vehicular Technology Conference (VTC)}, Sept 2015, pp. 1--5.

\bibitem{junil2016}
J.~Choi, V.~Va, N.~Gonzalez-Prelcic, R.~Daniels, C.~R. Bhat, and R.~W. Heath,
  ``Millimeter-wave vehicular communication to support massive automotive
  sensing,'' \emph{IEEE Communications Magazine}, vol.~54, no.~12, pp.
  160--167, December 2016.

\bibitem{refOfMatlab}
C.~Karnfelt, A.~Paden, A.~Bazzi, G.~E.~H. Shhade, M.~Abbas, and T.~Chonavel,
  ``77 {GHz} {ACC} radar simulation platform,'' in \emph{9th International
  Conference on Intelligent Transport Systems Telecommunications (ITST)}, Oct
  2009, pp. 209--214.

\bibitem{skolnik}
M.~I. {Skolnik}, \emph{{Radar Handbook}}, 3rd~ed.\hskip 1em plus 0.5em minus
  0.4em\relax New York: McGraw-Hill Education, 2008.

\bibitem{CohabitationCOM}
H.~Hellsten and P.~Dammert, ``Short range radar cohabitation: Extension to
  integrated communication,'' Saab Surveillance Report, 2017.

\bibitem{Lee2016}
S.~Lee, S.~Kang, S.~C. Kim, and J.~E. Lee, ``Radar cross section measurement
  with 77 {GHz} automotive {FMCW} radar,'' in \emph{{IEEE} 27th Annual
  International Symposium on Personal, Indoor, and Mobile Radio Communications
  (PIMRC)}, Sept 2016, pp. 1--6.

\end{thebibliography}

\end{document}